\documentclass{ws-procs11x85}

\usepackage{balance}

\newcommand{\gapprox}{\stackrel{>}{_{\sim}}}
\newcommand{\lapprox}{\stackrel{<}{_{\sim}}}

\makeindex
\begin{document}

\title{DEEP INELASTIC LEPTON-NUCLEON SCATTERING AT HERA}

\author{P. NEWMAN}

\address{School of Physics and Astronomy, University of Birmingham,
B15 2TT, UK\\E-mail: prn@hep.ph.bham.ac.uk}

\twocolumn[\maketitle\abstract{Data from the HERA collider experiments, 
H1 and ZEUS, have been fundamental to the rapid recent development of our 
understanding of the partonic composition of the proton and of QCD. 
This report focuses on inclusive measurements of neutral and charged
current cross sections at HERA, using the full available data taken to date. 
The present precision on the proton parton densities and the further
requirements for future measurements at the Tevatron and LHC are explored. 
Emphasis is also placed on the region of very low Bjorken-$x$ and $Q^2$. 
In this region, the `confinement' transition takes place from partons to 
hadrons as the relevant degrees of freedom and novel or exotic QCD
effects associated with large parton densities are most likely to be observed.
Finally, prospects for the second phase of HERA running are discussed.}]

\baselineskip=13.07pt
\section{The HERA Collider Experiments}

The presence of a point-like probe together with only one initial state 
hadron makes deep inelastic lepton nucleon scattering (DIS) the ideal
environment in which 
to study the quantum chromodynamics (QCD) of hadronic interactions
and to constrain the parton densities of the proton. 
In the years 1992-2000, the HERA collider experiments, H1 and ZEUS, collected
$ep$ data at electron 
beam energies of $27.5 \ {\rm GeV}$ and proton energies
of $820 \ {\rm GeV}$ and $920 \ {\rm GeV}$, corresponding to 
$ep$ center-of-mass energies
in excess of $300 \ {\rm GeV}$. The data were split between around 
$100 \ {\rm pb^{-1}}$ of $e^+ p$ collisions and $15 \ {\rm pb^{-1}}$ of
$e^- p$ collisions. 
With the extensions in accessible kinematic phase space
afforded by the large center-of-mass energy and the precise electron and 
hadron reconstruction in the experiments over a wide rapidity range, these
data have been used to gain new insights into many aspects of
$ep$ collisions. 

This article focuses mainly on measurements
of inclusive $ep$ cross sections in both neutral 
current (NC, $ep \rightarrow eX$)
and charged current 
(CC, $ep \rightarrow \nu X$) reactions throughout the available
phase space. In most cases, the full available data from the first phase 
of HERA running are used. At relatively large momentum transfers, the
inclusive cross sections 
yield important information on the parton densities of the proton. At small
momentum transfers, they can be used to search for novel effects in
quantum chromodynamics (QCD) associated with the large parton densities 
observed at the
previously unexplored low momentum fractions of the struck quark. 

Less inclusive measurements concentrating on other aspects of $ep$
scattering are described elsewhere in these proceedings.
The precision QCD tests and studies of the QCD evolution of parton cascades 
that have been possible through jet measurements are covered by Hirosky\rlap.\,\cite{hirosky}
Measurements of the structure of diffractive exchanges and insights into
the formation of rapidity gaps in hadronic interactions 
are discussed by Yamazaki\rlap.\,\cite{yamazaki}
Searches for new physics at HERA, at the highest $\surd s$
ever accessed in a collider with an initial state lepton, are covered 
by Perez\rlap.\,\cite{perez}

\section{Neutral and Charged Current DIS at Large {\boldmath $Q^2$}}
\label{highq2}

\begin{figure}
\center
\psfig{figure=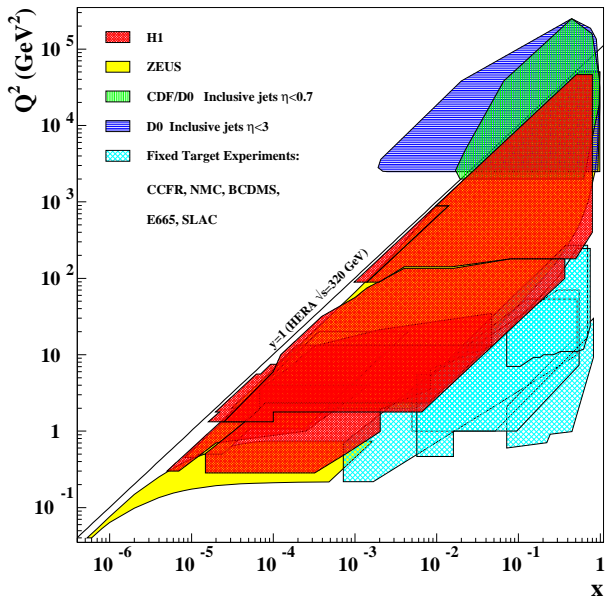,width=8.0truecm}
\caption[]{Kinematic plane in $x$ and $Q^2$ covered by inclusive HERA 
and fixed-target DIS measurements. The line at inelasticity
$y = 1$ represents the kinematic limit. 
The overlap region with recent jet measurements from the Tevatron\cite{tevjets}
is also illustrated.}
\label{kinplane}
\end{figure}

The kinematics of inclusive DIS are usually 
described by the variables $Q^2$, the
modulus of the squared four-momentum transfer carried by the exchanged 
electroweak gauge boson, and $x$, the  
fraction of the proton's longitudinal momentum carried by the quark that
couples to the exchanged boson. Figure~\ref{kinplane} illustrates the 
kinematic regions in which inclusive measurements have been made thus far
at HERA. 

The NC process takes place via the exchange of virtual photon and $Z^0$
propagators. The cross section can be expressed in the form
\begin{eqnarray}
  \label{ncform}
  \frac{{\rm d} \sigma^{\rm NC}}{{\rm d} x {\rm d} Q^2}
  = 2 \pi \alpha_{\rm em}^2 \cdot \left( \frac{1}{Q^2} \right)^2
  \cdot \frac{Y_+}{x} \cdot \tilde{\sigma}_{\rm NC} \ ,
\end{eqnarray}
where the term $\alpha_{\rm em}^2$ expresses the dominance of photon exchange
over most of the phase space, $1 / Q^4$ is the photon propagator term and
the reduced cross section
$\tilde{\sigma}_{\rm NC}$ contains helicity factors, weak terms
due to $Z^0$ exchange and structure functions related to the parton densities
of the proton. The variables $Y_\pm = 1 \pm (1 - y)^2$, dependent on the
inelasticity, $y$,
express the helicity dependence of the electroweak 
interaction (see also Eq.~(\ref{ncsf})).

The CC process is purely due to weak interactions. The cross
section can be expressed as
\begin{eqnarray}
  \label{ccform}
  \frac{{\rm d} \sigma^{\rm CC}}{{\rm d} x {\rm d} Q^2}
  = \frac{G_F^2 M_W^4}{2 \pi} \cdot \left( \frac{1}{Q^2 + M_W^2} \right)^2 
  \cdot \frac{1}{x} \cdot \tilde{\sigma}_{\rm CC} \ ,
\end{eqnarray}
where the coupling and propagator terms are specific to $W$ boson exchange
and the reduced cross section
term $\tilde{\sigma}_{\rm CC}$ contains the helicity factors and
structure functions.

\begin{figure}
\center
\psfig{figure=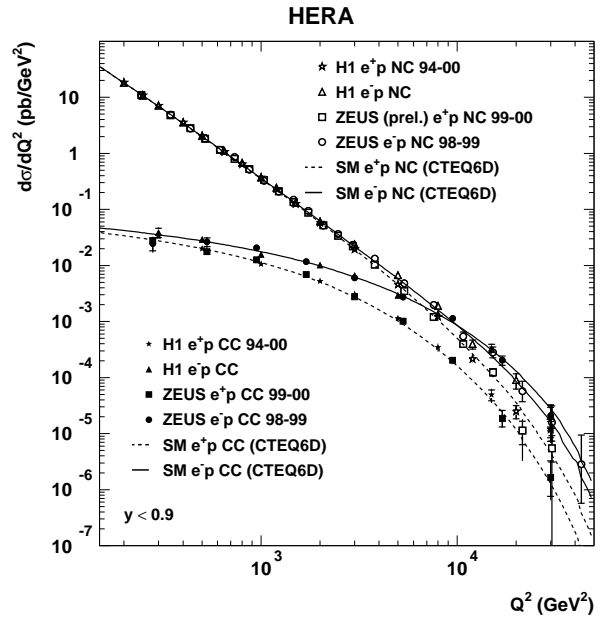,width=8.0truecm}
\caption[]{Single differential cross sections for charged and neutral
current interactions in $e^\pm p$ collisions, as measured by H1 and
ZEUS. The data are compared with the predictions of a recent global QCD
fit\rlap.\,\cite{cteq6}}
\label{ew}
\end{figure}

Figure~\ref{ew} shows the single differential cross sections measured
by H1 and ZEUS for charged and neutral current $e^\pm p$ scattering with
$Q^2 > 200 \ {\rm GeV^2}$\rlap.\,\cite{H1e-}$^-$\cite{ZEUSe+NC} 
For $Q^2 \ll M_W^2$, the NC cross section dominates heavily due to the
differences between the propagator terms in Eqs.~(\ref{ncform})
and~(\ref{ccform}). 
For $Q^2 \gapprox M_W^2$, the cross sections for NC and CC processes become
comparable, providing an illustration of electroweak unification with
space-like gauge bosons. The remaining differences between the NC and CC 
cross sections in this large $Q^2$ region and
the differences between the $e^+ p$ and $e^- p$ cross sections 
can be understood from the structure of the reduced cross sections
$\tilde{\sigma}_{\rm NC}$ and $\tilde{\sigma}_{\rm CC}$
(see Secs.~\ref{nc} and~\ref{cc}). 

\begin{figure}[h]
\center
\psfig{figure=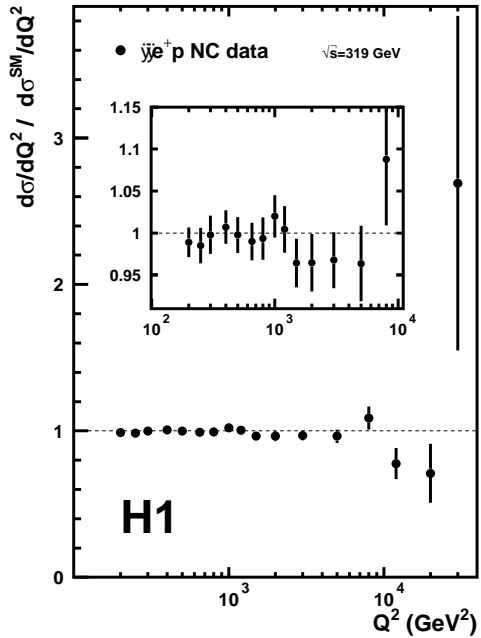,width=0.4\textwidth}
\caption[]{The ratio of H1 data to theoretical prediction for the single
differential NC cross section 
${\rm d} \sigma^{\rm NC} / {\rm d} Q^2$ in $e^+ p$
scattering. The theoretical predictions use the CTEQ5D\cite{cteq5}
parton densities, which were obtained using only a small fraction
of the HERA data.}
\label{ciplot}
\end{figure}

At the largest $Q^2$, the data are sensitive to possible new physics beyond
the Standard Model, for example due to quark compositeness. 
The data have been analyzed in the framework of possible contact 
interactions\rlap,\,\cite{ci} by comparing the measured
cross sections with predictions based on parton densities constrained mainly
by lower $Q^2$ and non-HERA data. 
An example of such a study is shown for NC $e^+ p$
interactions in Fig.~\ref{ciplot}. 
There is good agreement between the data and
the predictions based on the CTEQ5D\cite{cteq5} parton densities up to
the highest values of $Q^2 \sim 30\,000 \ {\rm GeV^2}$. As a result, quark
or electron substructure is excluded down to scales of
$\sim 1.0 \times 10^{-18} \ {\rm m}$.

\section{Parton Densities}

\subsection{Current Precision on Parton Densities}

Measuring and understanding the parton densities of the proton over as
wide a kinematic range as possible is a central aim of HERA analysis. 
This is important in its own right as a means of improving our understanding
of QCD. It is also crucial for precision measurements and the understanding
of backgrounds to searches for new physics at 
the Tevatron and LHC. As can be seen from 
consideration of the parton kinematics of $pp$
scattering\rlap,\,\cite{mrstlhc} 
accurate knowledge of the quark and gluon densities
over a wide range of $x$ is necessary to make precise Standard Model 
calculations for processes such as weak
gauge boson, Higgs, or top quark
production over the necessary rapidity ranges at
these machines. 

\begin{figure*}[tb]
\psfig{figure=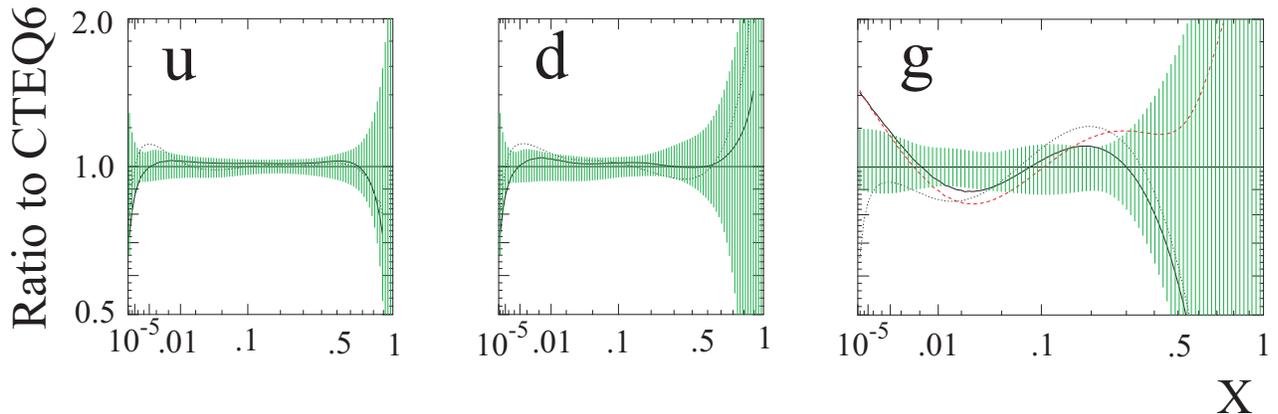,width=6.6truein}
\caption[]{
Estimates of the uncertainties on the proton parton densities at
$Q^2 = 10 \ {\rm GeV^2}$\rlap.\,\cite{cteq6} The bands show the
fractional uncertainty as a function of $x$.}
\label{cteq}
\end{figure*}

Figure~\ref{cteq}\cite{cteq6} shows estimates of the precision
with which the parton densities are currently known. Over the range 
$10^{-4} \lapprox x \lapprox 10^{-1}$ the combined experimental and 
theoretical uncertainties on the $u$ and $d$ 
quark densities are at the level of a few percent\rlap,\,\footnote{This level
of precision is only obtained after assumptions are made on the flavor
decomposition at low $x$, which have yet to be tested. For example, all
QCD fits assume that $\bar{u} - \bar{d} \rightarrow 0$ as $x \rightarrow 0$
and assumptions are necessary on the relative contributions from $s$ and $c$
quarks.}  
largely thanks to the low $x$ data
provided by HERA. The gluon density is somewhat more
poorly constrained over this
region. At large $x$ values, the uncertainties on all parton densities 
rapidly increase. This region is crucial for the understanding of
backgrounds to the production of any new particles near to threshold in
$pp$ scattering. 
Due to kinematic constraints (Fig.~\ref{kinplane}),
this high $x$ region can only be accessed at HERA at the highest $Q^2$,
where the cross section becomes small (Eq.~(\ref{ncform})). Large data 
samples are therefore needed to improve on the precision obtained from fixed-target
or other data in this region. 
In the following sections, the 
HERA data used to reach the levels of precision
shown in Fig.~\ref{cteq} are presented and discussed in particular in
terms of how improvements might be made at large $x$.

\subsection{Neutral Current Cross Sections}
\label{nc}

The reduced neutral current cross section
for $e^\pm p$ scattering, corrected for QED-radiative
effects, can be expressed as
\begin{eqnarray}
  \tilde{\sigma}_{\rm NC}^\pm = F_2 \ \mp \ \frac{Y_-}{Y_+} \ x F_3 \ 
- \ \frac{y^2}{Y_+} \ F_L \ ,
  \label{ncsf}
\end{eqnarray}
where $F_2$, $x F_3$ and $F_L$ are the generalized
unpolarized proton
structure functions. Extractions of $F_2$ and $x F_3$ are described in this
section. HERA $F_L$ data are discussed in Sec.~\ref{fl}.

The $F_2$ term is strongly dominant in most
of the measured phase space at HERA. After corrections for $Z^0$ exchange and 
interference between the photon and $Z^0$ contributions, the pure
electromagnetic structure function $F_2^{\rm em}$ can be extracted. 
In the
quark-parton model, this structure function can be decomposed as
\begin{eqnarray}
  F_2^{\rm em} (x, Q^2) = x \ \sum_q \ e_q^2 \ (q + \bar{q}) \ ,
\end{eqnarray}
where the sum runs over quark species $q$ of electric
charge $e_q$. 
$F_2^{\rm em}$ thus provides a squared-charge weighted sum
of all quark and antiquark densities.
Since $e_u^2 = 4 e_d^2$, it yields
a particularly strong constraint on the $u$ and $\bar{u}$ contributions. 

\begin{figure}[h]
\center
\psfig{figure=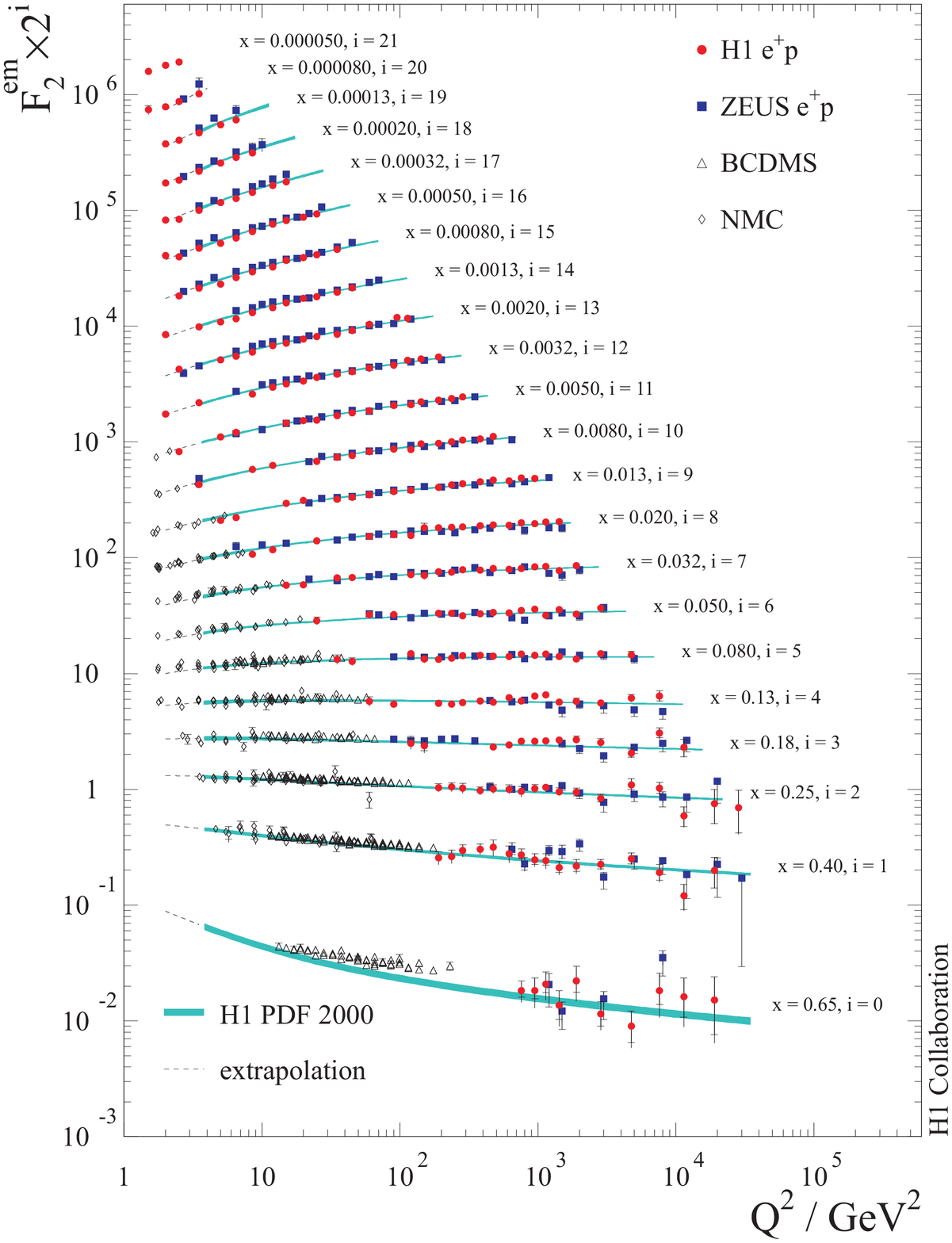,width=8truecm}
\caption[]{Summary of measurements of the structure function
$F_2^{\rm em}$ by H1 and ZEUS. The data are compared with the results of
a QCD fit to H1 NC and CC data only\rlap.\,\cite{H1e+}}
\label{F2}
\end{figure}

Figure~\ref{F2} shows a summary of the $F_2^{\rm em}$ data obtained
from $e^+ p$ scattering at HERA\rlap.\,\cite{H1e+,ZEUSe+NC} The structure
function is
measured over a huge kinematic range and the data are very well described 
over most of the range by QCD 
fits (see Sec.~(\ref{qcd}))\rlap.\,\footnote{A more complete discussion of the 
various QCD fits performed to DIS and other data and the estimates
of the corresponding uncertainties can be found
elsewhere in these proceedings\rlap.\,\cite{thorne}}
The precision reaches $2 - 3 \%$ in the bulk of the
phase space. However, in the region of the highest $x$, the precision
of the HERA data remains far from that of fixed-target experiments such as
BCDMS\cite{bcdms} and NMC\rlap,\,\cite{nmc} where the smaller center-of-mass
energy allows measurements at high $x$, but lower $Q^2$. At the highest $x$,
these fixed-target data are not well described by the QCD fits. It is 
highly desirable to obtain HERA measurements at high $x$ and intermediate
$Q^2$, where the potential problems of the fixed-target data (e.g.
higher twist
contributions and uncertainties in nuclear corrections) should be absent. 
This could be achieved with a reasonably large cross section by
running HERA with a reduced proton beam energy, as is planned as part of 
the second phase of HERA running. 

The $F_2^{\rm em}$
data also provide the best available constraints 
on the gluon density via the
deviations from Bjorken scaling, caused by gluon radiation. In leading order
of QCD, the gluon density can be obtained approximately from 
$\frac{\partial F_2^{\rm em} (x/2, Q^2)}{\partial \ln Q^2} 
\sim \alpha_s x g(x)$\rlap,\,\cite{prytz} 
such that the
strong positive scaling violations at low $x$ in Fig.~\ref{F2} are indicative
of a large and growing gluon density as $x$ becomes small. 

The structure function $x F_3$ arises due to $Z^0$ exchange. 
In the HERA phase space, the interference contribution
$x F_3^{\gamma Z}$ between the photon and $Z^0$ exchanges dominates, 
such that 
\begin{eqnarray}
x F_3 = -a_e \ \frac{\kappa Q^2}{Q^2 + M_Z^2} \ x F_3^{\gamma Z} + \Delta(Z^2)
\ ,
\end{eqnarray}
where in the quark-parton model, 
\begin{eqnarray}
x F_3^{\gamma Z} = 2 x \sum_q e_q a_q (q - \bar{q}) \ .
\end{eqnarray} 
Here, $a_e$ and $a_q$ are the axial couplings of the $Z^0$ to electrons
and quarks, respectively and
$\kappa^{-1} = 4 \frac{M_W^2}{M_Z^2} (1 - \frac{M_W^2}{M_Z^2})$ in the 
on-mass-shell scheme. Since $x F_3$ measures the difference between the
quark and antiquark densities, it is uniquely and model independently
sensitive to the non-singlet valence quark densities. 

\begin{figure}[h]
\center
\psfig{figure=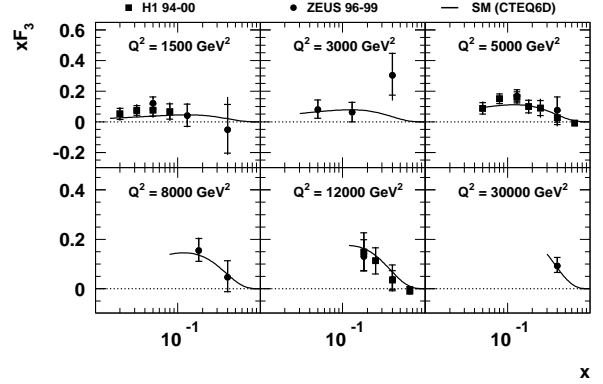,width=8.0truecm}
\caption[]{Data on $x F_3$ from HERA, based on around $100 \ {\rm pb^{-1}}$ of
$e^+ p$ data and $15 \ {\rm pb^{-1}}$ of $e^- p$ data,
compared with the predictions of
a recent global fit\rlap.\,\cite{cteq6}}
\label{xf3}
\end{figure}

Since it contributes with opposite signs to $e^+ p$ and $e^- p$ scattering
(Eq.~(\ref{ncsf})), $x F_3$ can be extracted from the 
measured differences between the NC $e^+ p$ and $e^- p$ 
cross sections at large $Q^2$, shown in Fig.~\ref{ew}, 
according to
\begin{eqnarray}
  x F_3 = \frac{Y_+}{2 Y_-} \left(\tilde{\sigma}_{\rm NC}^- - 
\tilde{\sigma}_{\rm NC}^+ \right) \ .
\end{eqnarray}

Figure~\ref{xf3} shows the current status of $x F_3$ data from 
HERA\rlap.\,\cite{H1e+,ZEUSe-NC} The
data are well described by the predictions of QCD fits in which the valence
quark densities are principally constrained by the NC and CC HERA data and
measurements from elsewhere, 
rather than by the differences between the $e^\pm p$ NC cross sections.
These measurements thus provide a test of the procedures used and
valence densities obtained in QCD fits. The large increases in luminosity 
expected at HERA-II are required to further exploit this observable. 

\subsection{Charged Current Cross Sections}
\label{cc}

In contrast to the NC measurements, where many millions of events are
available for analysis, the CC samples collected so far at HERA consist of
only around 1500 events ($e^+ p$) and 700 events ($e^- p$) per experiment.
The statistical uncertainties are
correspondingly larger than in the NC case, amounting to typically $6 \%$
for double differential measurements. 
Charged current cross section measurements do, however,
provide important complementary information for the extraction of parton 
densities, since they are sensitive to particular quark flavors with the
correct charges to couple to the exchanged $W$ boson. 

As can be
seen from Fig.~\ref{ew}, the $e^- p$ CC cross section is significantly
larger than the $e^+ p$ cross section throughout the measured kinematic
range. This is mainly because the $e^- p$ cross section is dominated by the
reaction $e^- u \rightarrow \nu_e d$, whereas the dominant $e^+ p$ process
is $e^+ d \rightarrow \bar{\nu}_e u$, the $u$ density being 
the larger in the
high $x$ region where measurements can be made. In the quark parton model, the
charged current reduced cross sections take the form
\begin{eqnarray}
  \tilde{\sigma}^-_{\rm CC} & = & x (u + c) \ + \ (1 -y)^2 \ x 
(\bar{d} + \bar{s}) \, \\
  \tilde{\sigma}^+_{\rm CC} & = & x (\bar{u} + \bar{c}) \ + \ 
(1 -y)^2 \ x (d + s) \ ,
\end{eqnarray}
where the helicity factor $(1 - y)^2$ implies a further kinematic suppression
to the term involving the $d$ density in the $e^+ p$ case. 

\begin{figure}[h]
\center
\psfig{figure=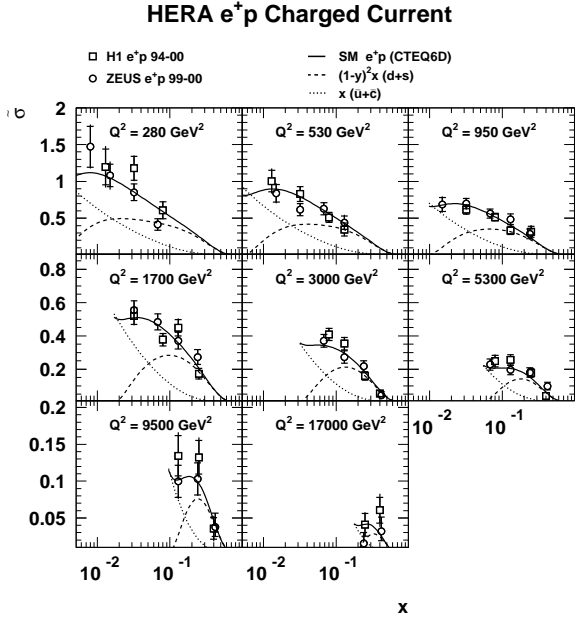,width=0.5\textwidth}
\caption[]{Double differential measurements of the CC $e^+ p$ cross section from
HERA. The data are compared with the predictions of a recent global 
fit\rlap,\,\cite{cteq6} which are broken down into contributions
from $d$-type and $\bar{u}$-type quarks.}
\label{cce+}
\end{figure}

The $e^- p$ CC cross section provides a complementary constraint to 
$F_2^{\rm em}$ on the $u$ density at high $x$\rlap.\,\cite{H1e-,ZEUSe-CC}
Despite the unfavorable helicity and smaller cross section, the $e^- p$
CC data yield the most direct available constraint on the $d$ density
at large $x$ from HERA. 
This is illustrated in Fig.~\ref{cce+}, which shows the most
recent measurements from H1\cite{H1e+} and ZEUS\rlap.\,\cite{ZEUSe+CC} The data
are compared with the predictions of a global QCD fit\rlap,\,\cite{cteq6} 
in which the $u$ and $d$ densities at large $x$ are constrained mainly
by precise fixed-target muon scattering data 
from protons and deuterons. The calculation
from the fit is broken
down into the contributions from scattering from $d$-type and $\bar{u}$-type 
quarks.

The CC measurements should improve considerably with HERA-II
data. However, even with $1 \ {\rm fb^{-1}}$, 
the uncertainties will remain large
in the 
important region $x \gapprox 0.5$. An alternative method of constraining
the $d$ density at large $x$ would be to run HERA with 
deuterons\cite{H1letter} and, using isospin symmetry, to unfold the $d / u$ 
ratio. Running with deuterons at $920 \ {\rm GeV}$ would also naturally
reduce the beam energy per nucleon, such that the benefits of the larger
cross section at large $x$ and intermediate $Q^2$ could be exploited. 

\begin{figure}[h]
\center
\psfig{figure=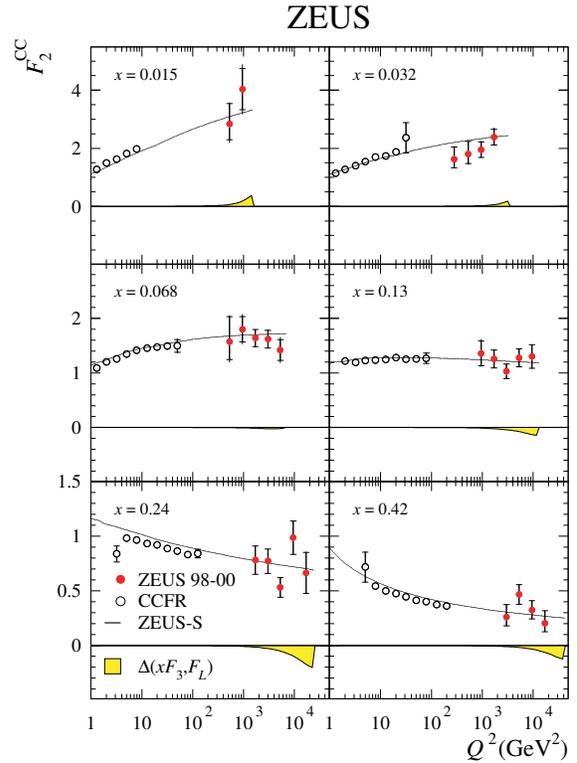,width=8.0truecm}
\caption[]{The flavor singlet CC structure function $F_2^{\rm CC}$ as obtained
by ZEUS and CCFR. The data are compared with the predictions of a QCD fit
to DIS data\rlap,\,\cite{zeusfit} which does not include the
CC data shown. The corrections necessary 
for the $x F_3^{\rm CC}$ and $F_L^{\rm CC}$ terms are also illustrated.}
\label{f2charged}
\end{figure}

The ZEUS collaboration\cite{ZEUSe+CC} have recently used their charged current
data to extract a flavor singlet CC structure function, 
\begin{eqnarray}
  F_2^{\rm CC} = \frac{2}{Y_+} \left( \tilde{\sigma}_{\rm CC}^+ 
+ \tilde{\sigma}_{\rm CC}^- \right) + \Delta(xF_3^{\rm CC}, F_L^{\rm CC}) \ .
\end{eqnarray}
This structure function is shown in Fig.~\ref{f2charged}, where 
the effects of the small $x F_3^{\rm CC}$ and 
$F_L^{\rm CC}$ correction
terms are also illustrated. The $F_2^{\rm CC}$ results
are compared with 
precise fixed-target neutrino data from CCFR\rlap.\,\cite{ccfr}
Viewing the fixed-target and HERA measurements together, the data span
more than four orders of magnitude in $Q^2$ and the influence of gluon
radiation on the CC process is clearly visible from the scaling violations. 
The ZEUS results  
are well described by the predictions of a QCD fit to various
DIS data\rlap.\,\cite{zeusfit}

\subsection{Parton Density Extractions}
\label{qcd}

The H1\cite{H1e+,H1alphas} and ZEUS\cite{zeusfit}
collaborations, along with various other groups\rlap,\,\cite{cteq6,mrst2001,alhekin}
have performed QCD fits to extract parton densities using various
combinations of HERA and other 
data. The fits are based on the evolution of the parton densities with
$Q^2$ using the DGLAP equations\cite{dglap} in Next-to-Leading Order 
(NLO)\rlap.\,\cite{dglapnlo} 

\begin{figure}
\center
\psfig{figure=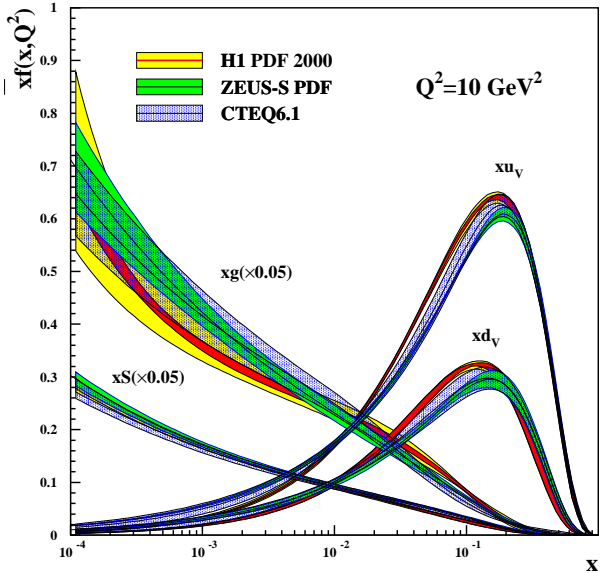,width=8.0truecm}
\caption[]{Parton densities as extracted by H1\rlap,\,\cite{H1e+} 
ZEUS\cite{zeusfit} and 
CTEQ\rlap.\,\cite{cteq6} The sea quark
and gluon densities are reduced by a factor of 20 for visibility.}
\vspace{-0.4cm}
\label{partons}
\end{figure}

With the latest HERA NC and CC data,
it is now possible to extract the full set of flavor-separated parton
densities from HERA data alone, provided assumptions are made on 
the validity
of DGLAP evolution throughout the fitted phase space and the
quark flavor decomposition at low $x$. Figure~\ref{partons} shows the results
for the valence densities, the sum of all sea quarks and the gluon density,
from H1 NC and CC data only\rlap,\,\cite{H1e+} 
ZEUS NC data together with other fixed-target DIS 
experiments\rlap,\,\cite{zeusfit} and CTEQ\rlap,\,\cite{cteq6} who perform a global fit 
to many DIS and other data sets. 
The agreement between the different extractions is reasonable, 
though there 
are differences between the H1 and ZEUS 
valence densities that go beyond the quoted error
bands. This is perhaps not surprising, given the
very different sources of information that are used to constrain the valence
densities in the two fits. H1 use the limited sensitivity to $W$ and $Z$
exchange effects in the HERA data to separate the valence and sea densities,
whereas ZEUS rely mainly on $x F_3$ data from fixed-target $\nu Fe$ and 
$\bar{\nu} Fe$ scattering\rlap.\,\cite{ccfrfe} The shapes of the gluon densities are 
also rather different. This arises from several sources, including
different parameterizations
of the parton densities at the starting scale for QCD evolution and
different treatments of heavy quark evolution\rlap.\,\footnote{The gluon densities
are thus defined in different schemes and strictly speaking cannot
be compared directly.}  
As $x \rightarrow 1$, the fixed-target data still give the best constraints. 
Deuteron running at HERA
is desirable to test the assumption, used in all fits,
that $\bar{d} - \bar{u} \rightarrow 0$ as $x \rightarrow 0$\rlap.\,\cite{H1letter}

\begin{figure}
\center
\psfig{figure=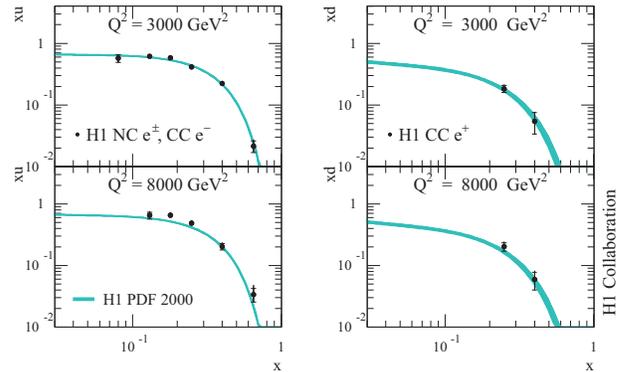,width=8.0truecm}
\caption[]{Results from the local extractions of parton densities by H1.
The $u$ density is obtained from NC $e^\pm$ and CC $e^-$ data points
for which it contributes in excess of $70 \%$ of the cross
section. The $d$ density is similarly extracted using CC $e^+$ data. The
data are compared with the results of a QCD fit to all H1 NC and CC
data\rlap.\,\cite{H1e+}}
\vspace{-0.4cm}
\label{local}
\end{figure}

Since the QCD fits rely on various assumptions, 
it is interesting to 
extract parton densities 
directly from
the data at fixed values of $x$ and $Q^2$,
in a manner that is relatively insensitive to these assumptions. 
As shown in Fig.~\ref{local}, the H1 collaboration has performed
such an extraction of the $u$ and $d$ densities at high $x$
using 
NC and CC data points for which the relevant parton contributes in excess of
$70 \%$ to the measured cross section according to the H1 QCD fit\rlap.\,\cite{H1e+} The
QCD fit is then used to correct for the remaining contributions. The 
resulting local
extractions of the $u$ and $d$ densities are in good agreement with the
predictions of the QCD fits using DGLAP evolution. 

\subsection{Tests of the Gluon Density}
\label{gluon}

The fits to inclusive data rely on the DGLAP QCD
evolution equations\cite{dglap,dglapnlo} to relate the scaling violations of 
$F_2^{\rm em}$ to the gluon density and require assumptions on the functional
form of the gluon density at the starting scale for QCD evolution. It is
important also to constrain the gluon density from other complementary
sources with different systematics, in order to test the 
overall consistency of the HERA data and the validity of
the assumptions of DGLAP evolution and QCD hard scattering factorization.
This has been done in several ways using hadronic final state data
at HERA. Measurements of dijet and charm production cross sections are
highly sensitive to the gluon density, since they proceed dominantly
via the boson-gluon fusion process $\gamma^* g \rightarrow q \bar{q}$,
the cross section for which is directly proportional to the gluon density
at leading order of QCD. The gluon density can thus be extracted in a manner
which is more sensitive to local variations. 

\begin{figure}[h]
\center
\psfig{figure=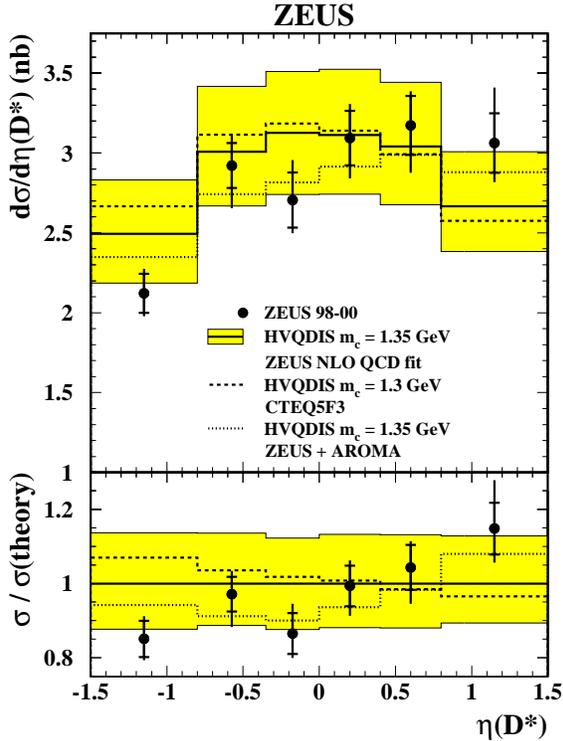,width=8.0truecm}
\caption[]{Cross section for $D^*$ meson production in DIS, differential
in pseudorapidity. The data are compared with the predictions of a NLO
QCD calculation based on parton densities from a fit to DIS 
data\rlap,\,\cite{zeusfit} together with an estimate of the
theoretical uncertainties. The effects of switching to the CTEQ5F3
parton densities\cite{cteq5}
or to Lund\cite{lund} rather than 
Petersen\cite{petersen} fragmentation 
(ZEUS + AROMA) are also indicated.}
\label{dstar}
\end{figure}

Examples of jet data that are sensitive to the gluon
density are given by Hirosky\rlap.\,\cite{hirosky} An example of recent 
charm data that constrain the gluon density 
can be found in Fig.~\ref{dstar}, which shows the cross section
for $D^*$ production in DIS as a function of pseudorapidity, as measured
by the ZEUS 
collaboration\rlap.\,\cite{zeusf2c} 
The data are compared with a theoretical prediction\cite{hvqdis}
based on NLO QCD matrix elements, interfaced to the gluon density from
a fit to inclusive DIS measurements\cite{zeusfit} and 
to fragmentation functions. The beautiful
agreement of the data and theory confirms the gluon density from scaling 
violations and the validity of the NLO DGLAP theory at the $10 \%$ level. 
The theoretical errors, dominated by the choice of the charm quark 
mass $m_c$, 
are larger than the uncertainties on the data. Comparing the predictions
using the ZEUS and CTEQ\cite{cteq5}
parton densities shows that the data can be used to
improve the constraints on 
the gluon density once the theoretical errors are better
controlled. 

By extrapolating the $D^*$ cross sections in the measured range of 
pseudorapidity and transverse momentum
to a charm production cross section integrated over the full phase
space, it is possible to extract $F_2^{c \bar{c}}$, the charm contribution
to the proton structure function $F_2$\rlap.\,\cite{zeusf2c,H1f2c} 
For $Q^2 \gg m_c^2$, 
such that the charm mass can be neglected, the ratio $F_2^{c \bar{c}} / F_2$
becomes close to $30 \%$ at low $x$, illustrating the importance of a 
proper treatment of the evolution of heavy quarks in any fit to
HERA data.

\section{The Low {\boldmath $x$} Region}

\subsection{Low $x$ Physics}

The region of low $x$, newly accessed at HERA, has been the subject of much
debate. The fast rise of the gluon density 
(Fig.~\ref{partons}) raises the question of whether unitarization effects
may become important\rlap.\,\footnote{It has been argued that the conventional
Froissart unitarity bound on hadronic total cross sections is not applicable 
to off-shell virtual photons\rlap.\,\cite{offshell}}
As the gluon density becomes large, the partons must
ultimately begin to interact through processes such as
gluon recombination ($gg \rightarrow g$)\rlap.\,\cite{glr} This would lead
to a taming of the low $x$ rise of $F_2$ and a breakdown of the DGLAP 
approximation.
 
A full perturbative QCD expansion gives rise to evolution of parton densities
with both $\ln Q^2$ and $\ln 1/x$. Standard DGLAP evolution is equivalent to
a resummation of leading $\ln Q^2$ terms, such that the struck quark 
originates from a parton cascade ordered in virtuality. At sufficiently
low $x$, evolution in $\ln 1/x$ must also become important, though it 
is not incorporated in the DGLAP approximation. Other approximations to
QCD evolution may then become more appropriate. Examples are BFKL 
evolution\rlap,\,\cite{bfkl} which resums the $\ln 1/x$ terms to all orders, or
CCFM evolution\rlap,\,\cite{ccfm} in which the partons are ordered in the angle
at which they are emitted. CCFM evolution is equivalent to BFKL evolution
for $x \rightarrow 0$, whilst limiting to the DGLAP equations at larger $x$. 

It has been suggested that the inclusion of BFKL effects improves the 
description of low $x$ inclusive measurements by QCD
fits\rlap,\,\cite{thornebfkl} though
no clear consensus exists on this question. There are also
hints from various
hadronic final state analyses that regions of phase space can be found at
HERA for which standard DGLAP evolution is insufficient and CCFM evolution
may provide a better description\rlap.\,\cite{fjets} 

Due to kinematic correlations (Fig.~\ref{kinplane}),
low $x$ values can only be accessed at low $Q^2$. The low $Q^2$ regime brings
its own complications, such as possible higher twist contributions and
the breakdown of convergence of perturbative
QCD as the strong coupling increases.
Around $Q^2 = 1 \ {\rm GeV^2}$, the ``confinement''
transition takes place, such that the partons of asymptotic freedom are
replaced by hadrons as the relevant degrees of freedom. 

\subsection{$F_2$ at Low $Q^2$}

\begin{figure}
\center
\psfig{figure=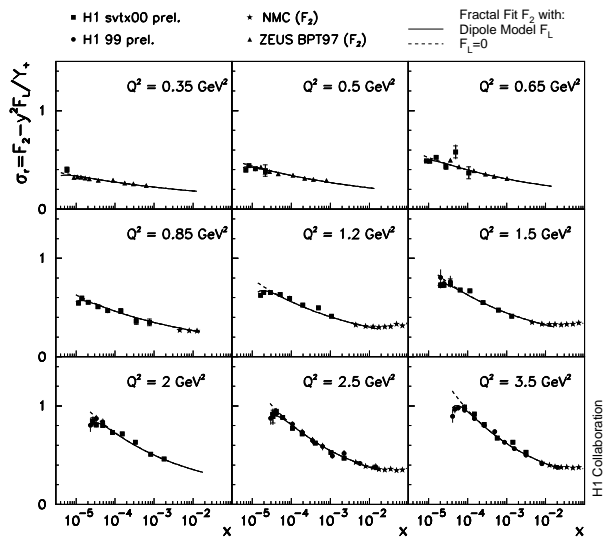,width=8.0truecm}
\caption[]{The reduced neutral current cross section $\tilde{\sigma}_{\rm NC}$
at low $Q^2$. The data are compared with the predictions of a `fractal'
model of proton structure\rlap.\,\cite{fractal}}
\label{sigmar}
\end{figure}

The data used in the measurements and QCD fits described in 
Sec.~\ref{highq2} cover the range $Q^2 \gapprox 3 \ {\rm GeV^2}$, where
perturbative QCD can reliably be used. The ZEUS collaboration has obtained
precise data (BPT97)
in the range $0.0045 < Q^2 < 0.65 \ {\rm GeV^2}$, using a
silicon strip tracking detector and an electromagnetic calorimeter very close
to the beampipe\rlap.\,\cite{bpt} Previously, the intermediate region, 
$0.65 < Q^2 \lapprox 3 \ {\rm GeV^2}$ 
has been 
only poorly explored at HERA, due to the acceptance limitations of the 
main detectors at small electron scattering angles. In order to improve this
acceptance, a short run was taken in 
the year 2000 with the $ep$ vertex shifted by
$70 \ {\rm cm}$ in the outgoing proton direction. The H1 collaboration has 
recently reported new inclusive NC measurements in the region 
$0.35 < Q^2 < 3.5 \ {\rm GeV^2}$, using these 
data\rlap.\,\cite{sv00} The resulting inclusive
cross section measurements are shown in the form of the reduced cross
section $\tilde{\sigma}_{\rm NC}$ in Fig.~\ref{sigmar}. The new data span 
the transition from a fast rise of the cross section with decreasing $x$ at
$Q^2 = 3.5 \ {\rm GeV^2}$ to a soft rise, similar to that observed in the
energy dependence of hadron-hadron total cross sections\rlap,\,\cite{dl} 
at $Q^2 = 0.35 \ {\rm GeV^2}$. At the lowest $x$ values, a decrease in the
cross section is observed due to the $F_L$ term in Eq.~(\ref{ncsf})
(see also Sec.~\ref{fl}).

\begin{figure}
\center
\psfig{figure=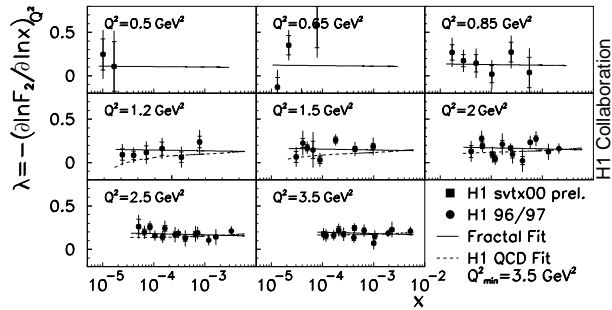,width=8.0truecm}
\caption[]{The logarithmic $x$ derivative of $F_2$ in the low $Q^2$ 
region. The data are compared with the predictions of a `fractal'
model of proton structure\rlap.\,\cite{fractal}}
\label{xderivs}
\end{figure}

In the double asymptotic limit\rlap,\,\cite{das} the DGLAP equations can be solved
with a solution whereby $F_2$ rises approximately as a power of $x$
as $x$ becomes small, such that $F_2 \sim x^{- \lambda}$. 
This feature is also predicted from the BFKL equations. Since 
unitarization effects would be expected to tame this growth, extracting
$\lambda$ has been suggested\cite{lambda} as a means of searching for
saturation effects. $\lambda$ corresponds to the logarithmic
$x$ derivative of $F_2$ at fixed $Q^2$,
\begin{eqnarray}
  \lambda(x, Q^2) = \left( \partial \ln F_2 / \partial \ln x \right)_{Q^2} \ ,
\end{eqnarray}
which
has been extracted locally from the differences between neighboring data
points in $x$ by the H1 collaboration\rlap.\,\cite{sv00,f2rise} 
The results in the kinematic region of the shifted vertex data are shown in 
Fig.~\ref{xderivs}. The data here and at larger $Q^2$
are consistent with no dependence of $\lambda$ on $x$ for fixed $Q^2$
and $x \lapprox 10^{-2}$, and thus with
a monotonic rise of $F_2$ as $x$ decreases with $Q^2$ fixed.
There is thus no evidence for any 
taming of this rise in inclusive electroproduction data from HERA. 

\begin{figure}
\center
\psfig{figure=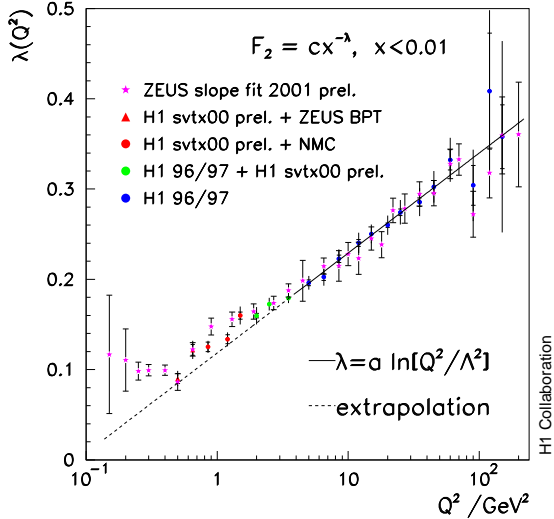,width=8.0truecm}
\caption{The results for $\lambda$ from fits to low $x$ data of the
form $F_2 \sim x^{- \lambda}$. The data are compared with a parameterization
in which $\lambda$ grows logarithmically with $Q^2$.}
\label{lambda}
\end{figure}

Since the logarithmic $x$ derivative is compatible with independence
of $Q^2$, 
the proton structure function at low $x$ can indeed be parameterized
as
\begin{eqnarray}
  F_2 = c(Q^2) \cdot x^{- \lambda(Q^2)} \ .
  \label{f2decomp}
\end{eqnarray}
H1 and ZEUS have both fitted their data to this 
form\rlap.\,\cite{sv00,f2rise,zeusregge} The results for 
$\lambda(Q^2)$ are shown in Fig.~\ref{lambda}. Two distinct regions seem
to be distinguished. For $Q^2 \gapprox 3 \ {\rm GeV^2}$, where partons are
the relevant degrees of freedom, $\lambda$ depends logarithmically on $Q^2$
and $c \sim 0.18$ is consistent with being constant. This behavior is well
reproduced by DGLAP-based QCD fits. In contrast, for 
$Q^2 \lapprox 1 \ {\rm GeV^2}$, there is evidence for a decrease in $c(Q^2)$
and for deviations of
$\lambda(Q^2)$ from the logarithmic dependence on $Q^2$
as it tends to the value of $0.08$ known to describe hadron-hadron\cite{dl}
and photoproduction\cite{sigmatot} total cross sections. In this region,
the description by DGLAP breaks down as the confinement transition takes
place on a distance scale of around $0.3 \ {\rm fm}$.  

\subsection{$F_L$ at Low $Q^2$}
\label{fl}

As can be seen from Fig.~\ref{sigmar}, the effects of $F_L$ are visible
in the inclusive 
reduced cross section at the lowest $x$, or highest $y$ values. 
In this region, the scattered electron energy becomes very small and
background from processes at $Q^2 \simeq 0$
in which a hadron is misidentified
as the scattered electron becomes large. The
H1 collaboration is able to make measurements for scattered electron energies
as low as $3 \ {\rm GeV}$ with the help
of drift chambers and a silicon tracking detector 
accompanying the backward calorimeter.
These detectors allow the event vertex to be reconstructed 
from the electron track at high $y$ and enable the suppression 
of photoproduction
background by ensuring that a track of the correct charge is linked to the 
electron candidate calorimeter cluster. 

$F_L$ is identically zero in lowest
order QCD, but acquires a non-zero value at ${\cal O} (\alpha_s)$ due to
gluon radiation. It is thus able to play a similar role to dijet and charm 
data  (Sec.~\ref{gluon}) 
in providing complementary information on the gluon 
density to that obtained
from the scaling violations of $F_2$ assuming DGLAP evolution. This is
particularly important at low $x$, where the $Q^2$ range of HERA
$F_2$ measurements is rather small (see Fig.~\ref{kinplane}), dijet and
charm measurements cannot be made due to kinematic restrictions and 
DGLAP evolution is most questionable. 
The sensitivity to $F_L$ at high
$y$, visible in Fig.~\ref{sigmar}, has been exploited
to determine $F_L$ in the crucial region
around $Q^2 = 1 \ {\rm GeV^2}$. 

\begin{figure}
\center
\psfig{figure=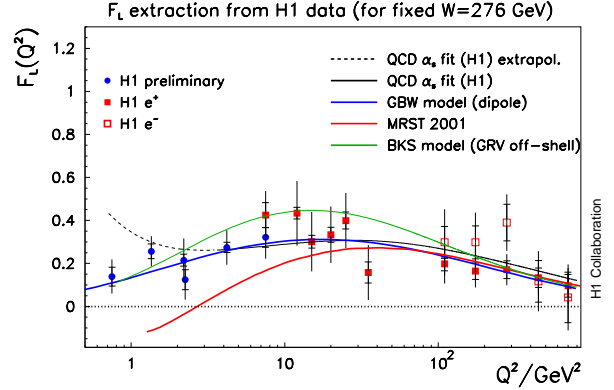,width=8.0truecm}
\caption[]{Summary of H1 $F_L$ determinations shown at fixed photon-proton
center-of-mass energy $W = 276 \ {\rm GeV}$. The data are compared with
the predictions of a QCD fit to H1 NC and CC data 
only\rlap,\,\cite{H1e+} a global QCD 
fit\rlap,\,\cite{mrst2001} a phenomenological dipole 
model\cite{kgbw} and a model based on unintegrated parton
densities and $k_T$ factorization\rlap.\,\cite{bks}.}
\label{flfig}
\end{figure}

The determination is made by fitting the reduced cross section to the form
\begin{eqnarray}
   \label{sigbreakdown}
   \tilde{\sigma}_{\rm NC} & = & F_2 - (y^2 / Y_+) \ F_L \\
                           & = & c(Q^2) \cdot x^{- \lambda(Q^2)} 
- (y^2 / Y_+) \ F_L \ ,
\end{eqnarray}
where the second equality follows from Eq.~(\ref{f2decomp}) and at
each $Q^2$ value, $c$, $\lambda$ and $F_L$ are free parameters. 
The results
of the extraction are insensitive to the assumptions on the behavior of
$F_2$ at the present level of accuracy. 
The shape of
the low $x$ turn-over of $\tilde{\sigma}_{\rm NC}$ is driven by the 
$y^2 / Y_+$ dependence, such that $F_L$ 
can only be extracted at a single point in $x$. 
The results are shown as a function of $Q^2$ 
in Fig.~\ref{flfig}, 
together with other $F_L$ extractions using similar
methods, spanning three orders of magnitude in 
$Q^2$\rlap.\,\cite{H1e-,H1e+,H1fl,mb99} 
The data are compared with a variety of
predictions based on DGLAP QCD fits\cite{H1alphas,mrst2001}
and other phenomenological approaches\rlap.\,\cite{kgbw,bks}
The data show that $F_L$ remains non-zero down to the lowest
$Q^2$ values measured and already distinguish between the
different models in the low $x$ region. 

Significant further progress in $F_L$ measurements 
at HERA can only be made by reducing
the proton beam energy, such that the $F_2$ and $F_L$ terms in 
Eq.~(\ref{sigbreakdown}) can be separated through measurements at the same
$x$ and $Q^2$, but different $y$. This would remove the need for assumptions
on the behavior of $F_2$ in the region where $F_L$ effects are present and
would allow measurements of the $x$ dependence, providing further important
discrimination between models.

\section{Future Prospects}

The HERA accelerator has recently restarted providing 
collisions, following a
shutdown during which it was upgraded to provide
a factor of around 
four increase in instantaneous luminosity. Spin rotators
and polarimeters have also been placed around the electron ring, so that
the effects of longitudinal polarisation of the electrons can be studied. 
In parallel, many upgrades have been made to the H1 and ZEUS detectors,
including improved silicon tracking, forward tracking and
track-based triggering. These improvements should improve 
the quality of data on charmed hadrons 
in particular and should extend the 
accessible phase
space for many final state measurements towards higher $x$. 

Over the next few years, the
aim is to collect $1 \ {\rm fb^{-1}}$ of data, representing a factor of
10 increase in statistics, equally shared between positron and electron
running with positive and negative lepton helicities. A run with
reduced proton beam energies in order to measure $F_L$ and access the high $x$,
intermediate $Q^2$ region is also planned. Further options for the running
of HERA, for example replacing the proton beam with deuterons or heavier ions
or polarising the proton beam in order to study nucleon spin 
at low $x$, do not currently form part of the future plans.

\section{Summary}

The data from the first phase of HERA running have now been fully analyzed
from the point of view of 
high $Q^2$ inclusive charged and neutral current cross sections.
The resulting data provide the best available constraints on the proton
quark and gluon densities in the region $10^{-4} < x < 10^{-1}$, crucial
for future experimentation at the Tevatron and LHC. Further
improvements at larger $x$ are possible in the future with 
higher luminosities and reduced proton
energy running. The data on hadronic final states 
are highly sensitive to the QCD of hadronic interactions and
complement
the inclusive measurements, providing tests of the QCD evolution equations
and competitive information on the gluon density. Here, improvements in
the precision of theoretical calculations are required in order to make 
significant further progress.
There have been considerable 
recent 
developments in understanding the region of low $x$ and $Q^2$ and testing
the range of validity of DGLAP evolution. With the HERA-II run just beginning,
the prospects are exciting for future measurements. 

\section*{Acknowledgments}

I would like to acknowledge the work of all members of the H1 and ZEUS
collaborations, which has led to the many impressive
results shown in this article. I
would also like to thank 
J. Butterworth, V. Chekelian, E. Elsen, T. Greenshaw,
B. Heinemann, V. Hudgson, M. Klein, 
T. La\v{s}tovi\v{c}ka, K. McFarland, D. Naples, E. Rizvi, R. Thorne and
R. Wallny for help with the preparation of this presentation and/or for
proof-reading this manuscript.

\clearpage
\balance
\twocolumn[
\section*{DISCUSSION}
]

\begin{description}

\item[Thomas Gehrmann] (Z\"{u}rich University):
Concerning the measurement of $F_2^{c \bar{c}}$, you mention that you obtain 
$F_2^{c \bar{c}}$ 
from an extrapolation of the charmed hadron spectra. Could you please
comment on how much $F_2^{c \bar{c}}$ is actually measurement, and how much is
extrapolation? And: how big is the error due to the extrapolation?

\item[Paul Newman:]
The extrapolations can be large. For the ZEUS measurement shown, they
vary from a factor of 5 at low $Q^2$ to a factor of 1.5 at high $Q^2$.
They are done using NLO QCD programs, the uncertainties are assessed 
and included in the quoted errors. They are not the dominant uncertainties, 
though assessing the full extrapolation errors is non-trivial.

\item[Rik Yoshida] (ZEUS spokesman, Argonne, paraphrased by PN):
While the systematic errors for the extrapolation of the $D^*$ cross section
to $F_2^{c \bar{c}}$ are evaluated, the extraction of $F_2^{c \bar{c}}$ is
necessarily a model dependent procedure. It is better to compare models 
of charm production with the data directly at the level of the measured 
differential cross sections. It is difficult to assess the correctness of 
models that only predict $F_2^{c \bar{c}}$ or the total charm cross section.

\item[Paul Newman:]
That is a very good point. The usefulness of $F_2^{c \bar{c}}$ lies mainly
in the illustration of the charm contribution to $F_2$ as a function of the
more familiar variables $x$ and $Q^2$. 
 
\item[Ikaros Bigi] (University of Notre Dame du Lac):
Do you see any evidence for intrinsic charm in the data,
or what is the status of this ancient concept?

\item[Paul Newman] (paraphrased and extended):
Our data are consistent with the charm component being entirely due to QCD
evolution and thus related to the gluon density. With the present data, there
is no need for an additional source. However, better data in the high $x$
region where intrinsic charm contributions have previously been discussed
are likely to become available at HERA-II, now that charm triggers and forward 
tracking have been improved.

\end{description}

\end{document}